\documentclass[aps,pre,groupedaddress]{revtex4}

\usepackage[dvipdfmx]{graphicx}
\usepackage{mathrsfs}
\usepackage{amsmath,amssymb}
\usepackage{hhline}
\usepackage{dcolumn}
\usepackage{bm}

\usepackage[dvips]{color}

\newcommand{\red}[1]{{}}

\newcommand{\be}{\begin{equation}}
\newcommand{\ee}{\end{equation}}
\newcommand{\bi}{\begin{itemize}}
\newcommand{\ei}{\end{itemize}}

\newcommand{\ea}{{\rm SN}}
\newcommand{\ma}{{\rm ON}}

\newcommand{\mn}{m_{\rm N}}
\newcommand{\ms}{m_{\rm S}}

\newcommand{\rC}{\rho_{0,{\rm c}}}
\newcommand{\rS}{\rho_{0,{\rm c}}^{\rm S}}
\newcommand{\sS}{S_{\rm S}}
\newcommand{\rD}{\rho_{0,{\rm c}}^{\rm SN}}
\newcommand{\sD}{S_{\rm SN}}
\newcommand{\rI}{\rho_{0,{\rm c}}^{\rm ON}}
\newcommand{\sI}{S_{\rm ON}}
\newcommand{\rDI}{\rho_{0,{\rm c}}^{\rm SN+ON}}
\newcommand{\sDI}{S_{\rm SN+ON}}
\newcommand{\rSD}{\rho_{0,{\rm c}}^{\rm S+SN}}
\newcommand{\sSD}{S_{\rm S+SN}}
\newcommand{\trho}{\tilde{\rho}_0}
\newcommand{\uS}{u_{\rm S}}

\newcommand{\uI}{u_{\rm ON}}
\newcommand{\uDI}{u_{\rm SN+ON}}
\newcommand{\uSD}{u_{\rm S+SN}}
\newcommand{\vSD}{v_{\rm S+SN}}
\newcommand{\wSD}{w_{\rm S+SN}}

\newcommand{\tS}{t_{\rm S}}
\newcommand{\tD}{t_{\rm SN}}
\newcommand{\tI}{t_{\rm ON}}
\newcommand{\tDF}{t_{\rm SN, FCC}}

\begin{document}

\title{Influence of initiators on the tipping point in the extended Watts model}
\date{\today}

\author{Takehisa Hasegawa}
\email{takehisa.hasegawa.sci@vc.ibaraki.ac.jp}
\affiliation{Graduate School of Science and Engineering, Ibaraki University, 2-1-1, Bunkyo, Mito, 310-8512, Japan}
\author{Shinji Nishioka}
\affiliation{Graduate School of Science and Engineering, Ibaraki University, 2-1-1, Bunkyo, Mito, 310-8512, Japan}

\begin{abstract}
In this paper, we study how the influence of initiators (seeds) affects the tipping point of information cascades in networks. 
We consider an extended version of the Watts model, in which each node is either active (i.e., having adopted an innovation) or inactive. 
In this extended model, the adoption threshold, defined as the fraction of active neighbors required for an inactive node to become active, depends on whether the node is a seed neighbor (i.e., connected to one or more initiators) or an ordinary node (i.e., not connected to any initiators).
Using the tree approximation on random graphs, we determine the tipping point, at which the fraction of active nodes in the final state increases discontinuously with an increasing seed fraction. 
The occurrence of a tipping point and the scale of cascades depend on two factors: whether a giant component of seed neighbors is formed when the seed fraction is large enough to trigger cascades among seed neighbors, and whether the giant component of ordinary nodes is maintained when newly activated nodes trigger further activations among ordinary nodes. 
The coexistence of two giant components suggests that a tipping point can appear twice. 
We present an example demonstrating the existence of two tipping points when there is a gap between the adoption thresholds of seed neighbors and ordinary nodes. 
Monte Carlo simulations clearly show that the first cascade, occurring at a small tipping point, occurs in the giant component of seed neighbors, while the second cascade, occurring at a larger tipping point, extends into the giant component of ordinary nodes.
\end{abstract}

\maketitle

\section{Introduction}

In recent years, there has been a growing interest among researchers in social contagion processes in networks, such as the spread of opinions and the adoption of innovations~\cite{jalili2017information, guilbeault2018complex}. 
In a pioneering study, Watts proposed a social contagion model (hereafter referred to as the Watts model) on a network, where each node representing an individual can be either active, indicating that it has adopted an innovation, or inactive~\cite{watts2002simple,watts2007influentials}. 
An inactive node changes its state to active by social influence if the fraction of active nodes among its neighbors exceeds a predetermined adoption threshold. 
The adoption threshold of a node represents its sensitivity to social influence: a lower adoption threshold signifies that fewer active neighbors are required for the node to become active, whereas a higher adoption threshold signifies that more active neighbors are required for the node to become active. 
In networks, the activation of a very small fraction of nodes can cause large cascades, where almost all nodes eventually become active if the mean adoption threshold is below a certain value~\cite{watts2002simple}.
The transition from a small cascade region, where cascades hardly spread, to a large cascade region, where cascades spread extensively, is often associated with a jump in the expected size of cascades.
Following the initial study by Watts~\cite{watts2002simple}, the relationship between the likelihood of large cascades and the structure of complex networks has been extensively investigated~\cite{ikeda2010cascade,hackett2011cascades,hackett2013cascades,gleeson2008cascades,payne2009information,dodds2009analysis,payne2011exact,galstyan2007cascading,nematzadeh2014optimal,curato2016optimal,yaugan2012analysis,zhuang2017clustering,unicomb2019reentrant}.
For example, studies on the Watts model in clustered networks~\cite{ikeda2010cascade,hackett2011cascades,hackett2013cascades} and correlated networks~\cite{gleeson2008cascades,payne2009information,dodds2009analysis,payne2011exact} have demonstrated that highly clustered structures and positively degree-correlated structures, respectively, facilitate the induction of large cascades.
Furthermore, some studies have proposed variants of the Watts model to model more complex contagion processes~\cite{dodds2004universal,dodds2005generalized,melnik2013multi,hurd2013watts,ruan2015kinetics,kobayashi2015trend,karsai2016local,lee2017social,huang2016contagion,wang2017heuristic,oh2018complex,chung2019susceptible}.

The occurrence of large cascades is dependent on the number of initiators (hereafter referred to as seeds), which are initially activated nodes, placed in the network~\cite{whitney2010dynamic}.
The critical fraction of seeds required to trigger large cascades is referred to as the {\it tipping point}~\cite{singh2013threshold,karampourniotis2015impact}.
Singh et al.~\cite{singh2013threshold} observed that when seeds are placed in a network using three strategies (random selection, target selection, and $k$-core selection), a discontinuous transition from the small cascade region to the large cascade region occurs at a tipping point. 
In addition, Karampourniotis et al.~\cite{karampourniotis2015impact} examined the Watts model, where the threshold of each node follows a normal distribution. 
They demonstrated that the expected size of cascades increases continuously rather than discontinuously as the seed fraction increases when the variance of the threshold distribution is large.  
The above studies do not distinguish between social influence originating from seeds and that from others. 
However, we can assume that seeds have a higher level of influence than others. 
Seeds may be technology enthusiasts or visionaries (innovators or early adopters in~\cite{moore2014crossing}) who are the first to adopt the innovation with great interest, or they may be chosen by marketers in viral marketing campaigns for attributes such as credibility and expertise. 
Nevertheless, the effect of the influence that seeds have on tipping points in the cascade dynamics of the Watts model has not yet been explored.

In this paper, we investigate the effect of the influence of seeds on the tipping point in cascade dynamics using the extended Watts model~\cite{nishioka2022cascading}.
In this extended model, nodes with connections to seeds (referred to as seed neighbors) have a different adoption threshold than ordinary nodes.
We apply the tree approximation to the extended Watts model on Erd\H{o}s--R\'{e}nyi random graphs (ERRGs) to determine the tipping points for different combinations of adoption thresholds for seed neighbors and ordinary nodes.
We observe that the occurrence of a tipping point and the scale of cascades are dependent on two factors: whether seed neighbors form a giant component (when the seed fraction is large enough to allow cascades of activations between seed neighbors), and whether ordinary nodes form a giant component (when activated nodes trigger further activations of ordinary nodes).
For example, if seed neighbors form a giant component, a large cascade can occur between them, even if ordinary nodes have such a high adoption threshold that almost none of them is ever activated.
In the case where a giant component of seed neighbors and that of ordinary nodes coexist, two tipping points can appear.
Our tree approximation demonstrates that for the extended Watts model with a gap between the adoption thresholds of seed neighbors and ordinary nodes, the active node fraction increases discontinuously at two different tipping points as the seed fraction increases. 
Monte Carlo simulations clearly show that the first cascade, occurring at a smaller tipping point, takes place in the giant component of seed neighbors, while the second cascade, occurring at a larger tipping point, extends into the giant component of ordinary nodes.

\section{Model}

This study employs the extended Watts model introduced in~\cite{nishioka2022cascading}.
Let us consider a network in which each node is in one of two states: active or inactive. 
A fraction $\rho_0$ of nodes are randomly selected as {\it seeds}, which are initially active. 
Among other nodes, which are initially inactive, {\it seed neighbors} are nodes that are connected to one or more seeds, while {\it ordinary nodes} are nodes that are not connected to any seeds~\footnote{In Ref.~\cite{nishioka2022cascading}, seed neighbors were called direct neighbors, while ordinary nodes were called indirect neighbors. However, these terms were confusing and have been corrected in this paper.}.
Whether seed neighbors and ordinary nodes become active depends on the predetermined adoption thresholds, $\theta_{\rm SN}$ and $\theta_{\rm ON}$, respectively.
A seed neighbor of degree $k$ changes its state to active if the proportion of $k$ adjacent nodes that are active exceeds a predetermined threshold $\theta_\ea$, that is, $m \ge k \theta_\ea$, where $m$ denotes the number of active neighbors. Similarly, an ordinary node of degree $k$ changes its state to active if the proportion of $k$ adjacent nodes that are active exceeds a predetermined threshold $\theta_\ma$, that is, $m \ge k \theta_\ma$.

We distinguish between the adoption thresholds of seed neighbors and ordinary nodes to assume $\theta_{\rm SN} < \theta_{\rm ON}$. 
Seed neighbors can be considered as individuals who are directly influenced by the initial information source, which may be the first to adopt the innovation with great interest as technology enthusiasts or visionaries. 
Their proximity to the source makes them more likely to be easily activated, reflecting their lower adoption threshold $\theta_{\rm SN}$. 
Ordinary nodes, on the other hand, represent individuals who are not directly connected to the initial information source. 
These nodes typically require a larger amount of influence to become activated due to their distance from the source, reflecting their higher adoption threshold $\theta_{\rm ON}$.

The cascade dynamics begins from an initial state in which seeds are active and non-seed nodes are inactive.
Once activated, nodes always remain active, and the updating process continues until no additional nodes become active. 
The fraction of active nodes in the final state is denoted by $\rho_\infty (\ge \rho_0)$, where $\rho_\infty$ describes the scale of the cascade: $\rho_\infty \approx \rho_0$ when cascades triggered by seeds hardly spread, and $\rho_\infty \approx 1$ when cascades triggered by seeds cause almost all nodes in the network to be active in the final state.

The tree approximation proposed by Gleeson and Cahalane~\cite{gleeson2007seed} describes the Watts model starting with an arbitrary seed fraction $\rho_0$.
Following \cite{gleeson2007seed}, a tree approximation was presented for the extended Watts model on random networks~\cite{nishioka2022cascading}.
For infinitely large and random networks with degree distribution $p_k$, the active node fraction $\rho_\infty$, which represents the probability that a randomly selected node is active in the final state, is determined as follows: 
\begin{eqnarray}
\rho_\infty
&=&
\rho_{0}+\left(1-\rho_{0}\right) \sum_{k=1}^{\infty} p_k\left(1-\rho_{0}\right)^{k} \sum_{\mn=0}^{k} \binom{k}{\mn} r_{\infty}^{\mn}\left(1-r_{\infty}\right)^{k-\mn} F_\ma\left(\frac{\mn}{k}\right) \label{eq:rhoinfty} \\
&&+\left(1-\rho_{0}\right) \sum_{k=1}^{\infty} p_k \sum_{\ms=1}^{k} \binom{k}{\ms} \rho_{0}^{\ms}\left(1-\rho_{0}\right)^{k-\ms} \sum_{\mn=0}^{k-\ms} \binom{k-\ms}{\mn} r_{\infty}^{\mn}\left(1-r_{\infty}\right)^{k-\ms-\mn }F_\ea\left(\frac{\ms+\mn}{k}\right), \nonumber
\end{eqnarray}
where $\ms$ denotes the number of adjacent seeds, $\mn$ denotes the number of active adjacent non-seed nodes (i.e., seed neighbors and ordinary nodes), $F_{\ea}(x)$ denotes the response function of seed neighbors, defined as
\be
F_{\rm \ea}(x)=\left\{\begin{array}{l}{1~~~~{\rm if} ~~x \ge \theta_{\rm \ea}} \\ {0~~~~{\rm otherwise}}\end{array}\right.,
\ee
$F_{\ma}(x)$ denotes the response function of ordinary nodes, defined as
\be
F_{\rm \ma}(x)=\left\{\begin{array}{l}{1~~~~{\rm if} ~~x \ge \theta_{\rm \ma}} \\ {0~~~~{\rm otherwise}}\end{array}\right.,
\ee
and $r_\infty$ denotes the probability that a randomly selected neighbor of a non-seed node is active in the final state.
Equation (\ref{eq:rhoinfty}) for $\rho_\infty$ is rewritten as follows:
\begin{eqnarray}
\rho_\infty
&=& \rho_0 + (1-\rho_0) \sum_{k=0}^{\infty} p_k \sum_{m=0}^k \binom{k}{m} ((1-\rho_0)(1-r_\infty))^{k-m} (\rho_0+(1-\rho_0)r_\infty)^m F_\ea \left(\frac{m}{k}\right) \nonumber \\
&&+(1-\rho_0)\sum_{k=0}^{\infty} p_k \left(1-\rho_{0}\right)^{k} \sum_{m=0}^{k} \binom{k}{m} r_\infty^{m}(1-r_\infty)^{k-m}\left(F_\ma\left(\frac{m}{k}\right) - F_\ea\left(\frac{m}{k}\right)\right),  \label{eq:rho}
\end{eqnarray}
and $r_\infty$ is given as the solution of the following equation:
\begin{eqnarray}
r_{\infty} 
&=& 
\sum_{k=0}^{\infty} q_k \sum_{m=0}^k \binom{k}{m} ((1-\rho_0)(1-r_\infty))^{k-m} (\rho_0+(1-\rho_0)r_\infty)^m F_\ea \left(\frac{m}{k+1}\right) \nonumber \\
&&+\sum_{k=0}^{\infty} q_k \left(1-\rho_{0}\right)^{k} \sum_{m=0}^{k} \binom{k}{m} r_\infty^{m}(1-r_\infty)^{k-m}\left(F_\ma\left(\frac{m}{k+1}\right) - F_\ea\left(\frac{m}{k+1}\right)\right), \label{eq:r}
\end{eqnarray}
where $q_k$ is the excess degree distribution, expressed as $q_k = (k+1)p_{k+1}/\langle k \rangle$, with $\langle k \rangle$ representing the average degree $\langle k \rangle=\sum_k k p_k$.

By evaluating $r_\infty$ using Eq.~(\ref{eq:r}) and substituting it into Eq.~(\ref{eq:rho}), we obtain the active node fraction $\rho_\infty$ of the extended Watts model with arbitrary parameters ($\rho_0,~\theta_\ea,~\theta_\ma$) on a given network. 
A previous study~\cite{nishioka2022cascading} examined the extended Watts model on ERRGs, whose degree distribution is $p_k = \langle k \rangle^k e^{-\langle k \rangle}/k!$, starting from a small seed fraction $\rho_0$.
It was found that increasing the influence of seeds considerably increases the probability of large cascades by making it easier for a small number of seed neighbors to become active. 
In Sec.~\ref{sec:results}, we employ the tree approximation for the extended Watts model on ERRGs starting with an arbitrary seed fraction to examine the effect of the influence of seeds on the tipping point, at which the active node fraction $\rho_\infty$ increases explosively.
We also perform Monte Carlo simulations for the extended Watts model on ERRGs to validate the tree approximation.
For all cases presented in Sec.~\ref{sec:results}, we find that our analytical results using the tree approximation are in good agreement with Monte Carlo simulation results.

\section{Results} \label{sec:results}

\begin{figure}[tb]
\centering
(a)
\includegraphics[width=7.cm]{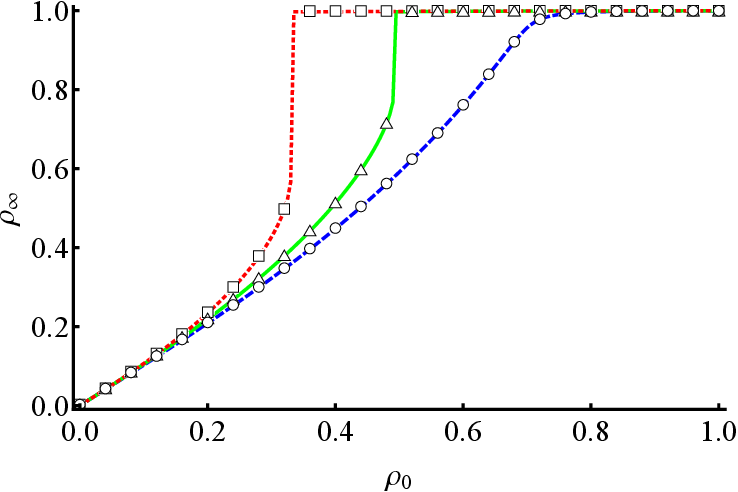}
(b)
\includegraphics[width=7.cm]{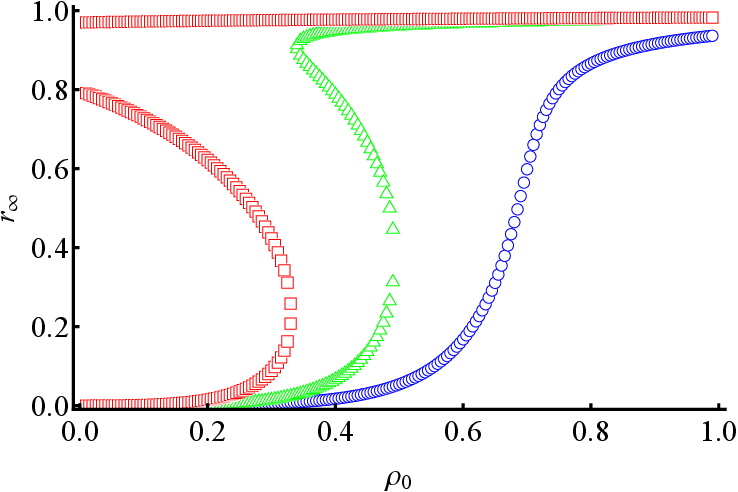}
\caption{
Results for the original Watts model on an ERRG with $\langle k\rangle = 6$: 
(a) active node fraction $\rho_\infty$ evaluated using Eq.~(\ref{eq:rho}) and Eq.~(\ref{eq:r}), and (b) fixed points $r_\infty$ obtained using Eq.~(\ref{eq:r}).
The results for $\theta=0.55$, $0.65$, and $0.75$ are represented by the red dotted line, green solid line, and blue dashed line, respectively, in (a), and by red squares, green triangles, and blue circles, respectively, in (b).
The symbols in (a) represent the simulation results for ERRGs with $10^5$ nodes, averaged over $10^3$ samples ($10^2$ graphs $\times$ $10$ initial states).
The theoretical lines are in good agreement with the simulation results. 
}
 \label{fig:originalWatts}
 \end{figure}

First, we briefly review the case of  $\theta_\ea=\theta_\ma=\theta$, where the extended Watts model corresponds to the original Watts model with threshold $\theta$.
Figure~\ref{fig:originalWatts}(a) plots the active node fraction $\rho_\infty$ as a function of the seed fraction $\rho_0$ for the Watts model on an ERRG with average degree $\langle k \rangle =6$.
For $\theta=0.55$ and $0.65$, the active node fraction increases continuously with $\rho_0 (<\rC)$,  jumping to $\rho_\infty \approx 1$ at the tipping point $\rho_0 = \rC(\theta)$.
A higher adoption threshold $\theta$ results in a tipping point $\rC(\theta)$ appearing at a larger seed fraction.
For large values of $\theta$, no discontinuous transitions are observed, as seen in the case of  $\theta=0.75$.
Figure~\ref{fig:originalWatts}(b) displays the fixed points $r_\infty$ of Eq.~(\ref{eq:r}) as a function of $\rho_0$.  
When $\theta$ is low, there are three fixed points for small $\rho_0$ and one fixed point for large $\rho_0$, signifying that a discontinuous transition occurs at some seed fraction. 
In contrast, when $\theta$ is high, there is always one fixed point regardless of $\rho_0$, signifying that there is no jump in $\rho_\infty$.
In general, the seed fraction required for activations to spread increases with an increase in $\theta$. 
However, cascades of activations are possible only on the connected components of non-seed nodes, which become smaller and split as the number of seeds increases. 
The disappearance of jumps in $\rho_\infty$ for large values of $\theta$ is attributed to the fact that the connected components of non-seed nodes are split into finite ones due to an excessive number of seeds before activations begin to spread.
For a large cascade to occur in the network, non-seed nodes must be sufficiently connected to form a {\it giant component}, occupying a finite fraction of nodes in the network, when the seed fraction becomes large enough to allow activations to propagate.

\begin{figure}[tb]
\centering
(a)
\includegraphics[width=7.cm]{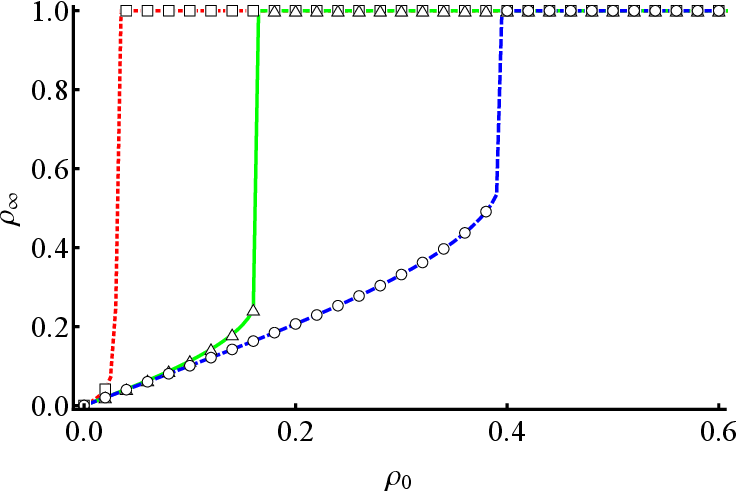}
(b)
\includegraphics[width=7.cm]{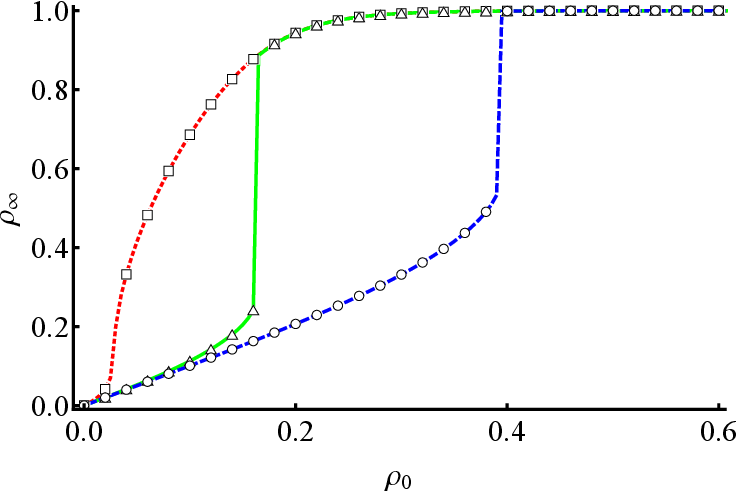}
\caption{
Active node fraction $\rho_\infty$ as a function of $\rho_0$ for the extended Watts model when the adoption threshold of ordinary nodes is set to (a) $\theta_\ma = 0.50$ and (b) $\theta_\ma = 0.90$.
We employed an ERRG with $\langle k \rangle = 10$ and evaluated $\rho_\infty$ using Eqs.~(\ref{eq:rho}) and~(\ref{eq:r}).
The red dotted line, green solid line, and blue dashed line represent the results when the adoption threshold of seed neighbors is $\theta_\ea = 0.2, 0.4$, and $0.6$, respectively.
The symbols represent the simulation results for ERRGs with $10^5$ nodes, averaged over $10^3$ samples ($10^2$ graphs $\times$ $10$ initial states).
The theoretical lines are in good agreement with the simulation results.
}
 \label{fig:extendedWatts}
 \end{figure}

Next, we consider the case of $\theta_\ea \neq \theta_\ma$.
Figures~\ref{fig:extendedWatts}(a) and (b) plot the active node fraction $\rho_\infty$ as a function of the seed fraction $\rho_0$ for the extended Watts model on an ERRG with average degree $\langle k \rangle=10$, where the adoption threshold of ordinary nodes is set to $\theta_\ma=0.5$ and $0.9$, respectively.
As illustrated in Fig.~\ref{fig:extendedWatts}(a), when $\theta_\ma$ is low (i.e., ordinary nodes can easily become active), then when seed neighbors are activated by seeds, further activations are triggered by active seed neighbors, resulting in an activation cascade of ordinary nodes.
In this case, the active node fraction jumps to $\rho_\infty \approx 1$ at the tipping point $\rC$.
In contrast, as illustrated in Fig.~\ref{fig:extendedWatts}(b), when $\theta_\ma$ is high (i.e., ordinary nodes do not easily become active), activation cascades may not occur between ordinary nodes even when most seed neighbors become active.
When $\theta_\ea=0.2$ and $0.4$, $\rho_\infty$ increases discontinuously at the tipping point but does not reach $\rho_\infty \approx 1$, and increases continuously as $\rho_0$ increases above the tipping point.
At the tipping point, ordinary nodes are rarely activated, and activations spread only over the giant component formed by seed neighbors.

\begin{figure}[tb]
\centering
(a)
\includegraphics[width=4cm]{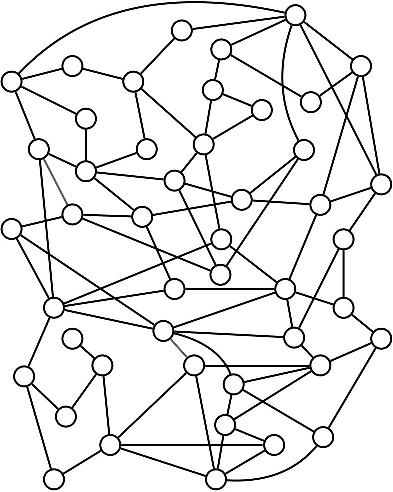}
(b)
\includegraphics[width=4cm]{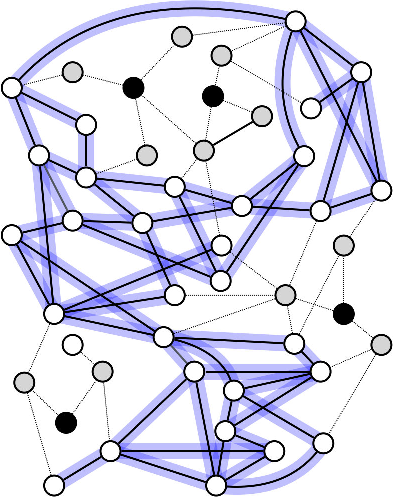}
(c)
\includegraphics[width=4cm]{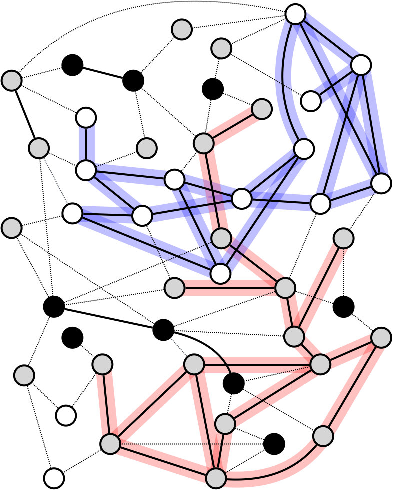}

(d)
\includegraphics[width=4cm]{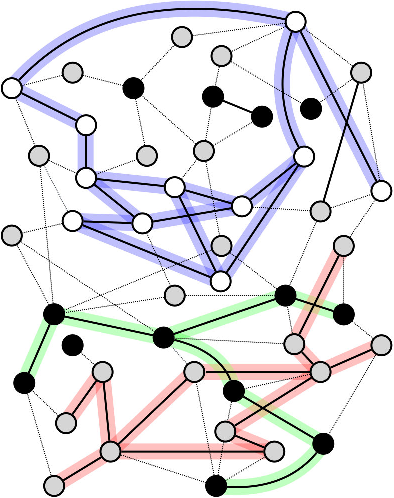}
(e)
\includegraphics[width=4cm]{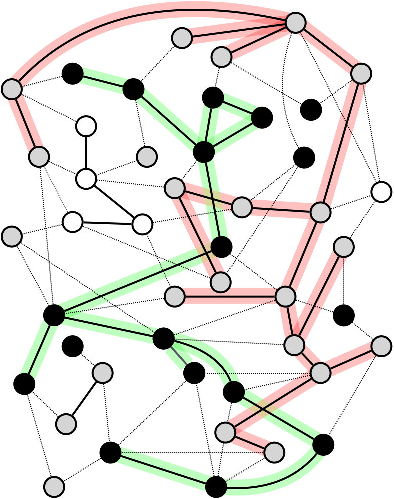}
(f)
\includegraphics[width=4cm]{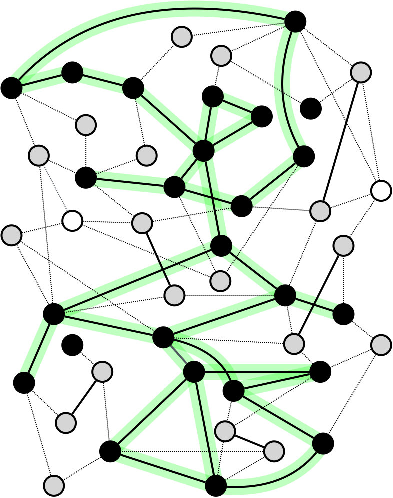}
\caption{
Example of how nodes of each type form a giant component at different seed fractions. 
The black circles, gray circles, and white circles represent seeds, seed neighbors, and ordinary nodes, respectively.
(a) A network before seed placement.
(b) When $\rho_0 < \rho_{0, {\rm c}}^{\rm SN}$ (where $\rho_{0, {\rm c}}^{\rm SN}$ is the smaller critical point), only the ordinary nodes form a giant component.
(c) When the seed fraction increases to $\rho_{0, {\rm c}}^{\rm SN} < \rho_0 < \rho_{0, {\rm c}}^{\rm S}$, both seed neighbors and ordinary nodes form their own giant components.
(d) When $\rho_{0, {\rm c}}^{\rm S} < \rho_0 < \rho_{0, {\rm c}}^{\rm ON}$, seeds also form their own giant component.
(e) When $\rho_{0, {\rm c}}^{\rm ON} < \rho_0 < \rho_{0, {\rm c}}^{\rm SN}$ (where $\rho_{0, {\rm c}}^{\rm SN}$ is the larger critical point), the giant ON-component is dismantled by seeds and seed neighbors, while seeds and seed neighbors form their own giant component.
(f) When $\rho_0 > \rho_{0, {\rm c}}^{\rm SN}$, the number of seeds becomes too large, and seed neighbors also lose their giant component. As a result, only the seeds form a giant component.
}
 \label{fig:typeofGC}
 \end{figure}

\begin{figure}[tb]
\centering
(a)
\includegraphics[width=7.cm]{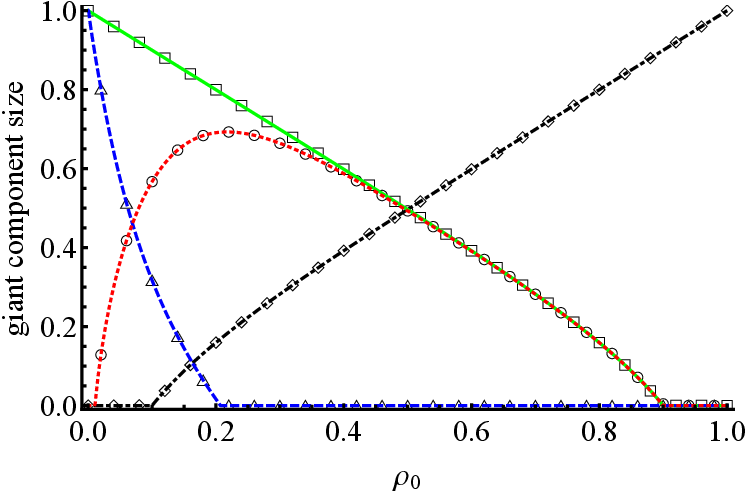}
(b)
\includegraphics[width=7.cm]{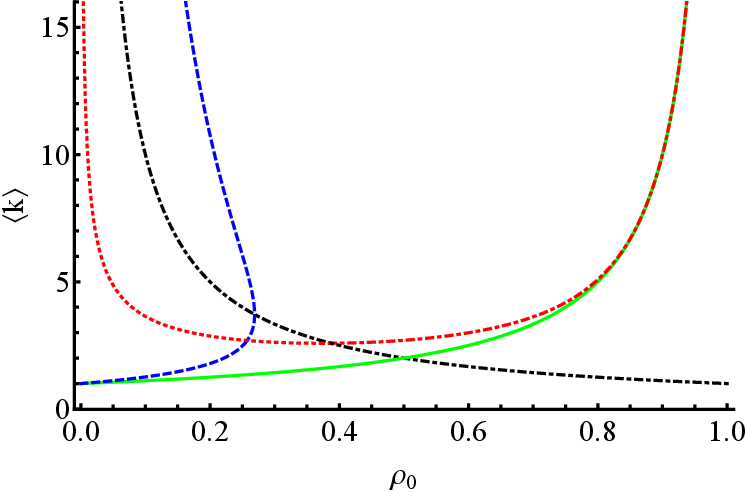}
\caption{
(a) Giant S-component size $\sS$ (black dot-dashed line), giant SN-component size $\sD$ (red dotted line), giant (SN+ON)-component size $\sDI$ (green solid line), and giant ON-component size $\sI$ (blue dashed line) as a function of $\rho_0$.
An ERRG with $\langle k\rangle=10$ was employed.
(b) Critical point $\rS$ for S-components (black dot-dashed line), $\rD$ for SN-components (red dotted line), $\rDI$ for (SN+ON)-components (green solid line), and $\rI$ for ON-components (blue dashed line) in the $(\rho_0, \langle k \rangle)$ plane.
The symbols in (a) represent the simulation results for ERRGs with $10^5$ nodes, averaged over $10^3$ graphs.
The theoretical lines are in good agreement with the simulation results. 
}
 \label{fig:component}
 \end{figure}

In the present model, it is natural to assume that the scale of cascades is related to whether nodes of each type form a giant component at different seed fractions.
We refer to the connected components of seeds, the connected components of seed neighbors, the connected components of ordinary nodes, and the connected components of seed neighbors and ordinary nodes as S-components, SN-components, ON-components, and (SN+ON)-components, respectively.
Whether each type of connected component forms its own giant component, which occupies a finite fraction of a network, depends on the seed fraction $\rho_0$ (Fig.~\ref{fig:typeofGC}).
Using generating functions, we derive the size of these components and their associated critical points (Appendix~\ref{sec:Appendix}).
Figure~\ref{fig:component}(a) presents the normalized giant S-component size $\sS$, giant SN-component size $\sD$, giant ON-component size $\sI$, and giant (SN+ON)-component size $\sDI$ of the ERRG with $\langle k \rangle =10$ as a function of the seed fraction $\rho_0$.
The giant S-component appears above a critical point $\rS = 0.1$ when $\rho_0$ increases ($\sS = 0$ for $\rho_0 \le \rS$ and $\sS >0$ for $\rho_0 > \rS$).
The giant ON-component and the giant (SN+ON)-component occupy most of the network when the seed fraction is small; however, their sizes decrease with $\rho_0$ and disappear at $\rI \approx 0.207$ and $\rDI = 0.9$, respectively ($\sI =0$  for $\rho_0 \ge \rI$ and $\sDI =0$ for $\rho_0 \ge \rDI$).
When $\rI < \rho_0 <\rDI$, where $\sI = 0$ but $\sDI>0$, the giant (SN+ON)-component is mostly composed of seed neighbors.

The giant SN-component behaves non-monotonically with an increasing seed fraction $\rho_0$. 
As $\rho_0$ increases from zero, the number of seed neighbors first increases, and the giant SN-component is formed (Fig.~\ref{fig:typeofGC}(c)) at a smaller critical point ($\rD \approx 0.011$).
However, as the seed fraction continues to increase, the connectivity between seed neighbors is obstructed by seeds.
As the seed fraction exceeds a larger critical point ($\rD \approx 0.900$), the giant SN-component disappears (Fig.~\ref{fig:typeofGC}(f)).
Thus, the giant SN-component exists only in the region where $\rho_0$ is between the two values of $\rD$\; ($\sD > 0$ if $0.011 \lessapprox \rho_0 \lessapprox 0.900$ and $\sD = 0$ otherwise).
Figure~\ref{fig:component}(b) plots the critical point for each type of giant component in the ($\rho_0$, $\langle k \rangle$) plane.
The figure indicates that when the average degree of the ERRG is $\langle k \rangle>\langle k \rangle_{\rm c} \approx 2.580$, each type of giant component behaves qualitatively the same as in Fig.~\ref{fig:component}(a). 
However, when $1 < \langle k \rangle<\langle k \rangle_{\rm c}$, seed neighbors never form a giant component.
In the absence of a giant SN-component, a jump in $\rho_\infty$ does not occur unless there is a cascade of activations between ordinary nodes.
However, when the giant SN-component is present, it is possible for a cascade to occur among seed neighbors, resulting in a jump in $\rho_\infty$ even if ordinary nodes are rarely activated.

\begin{figure}[tb]
\centering
(a)
\includegraphics[width=7.5cm]{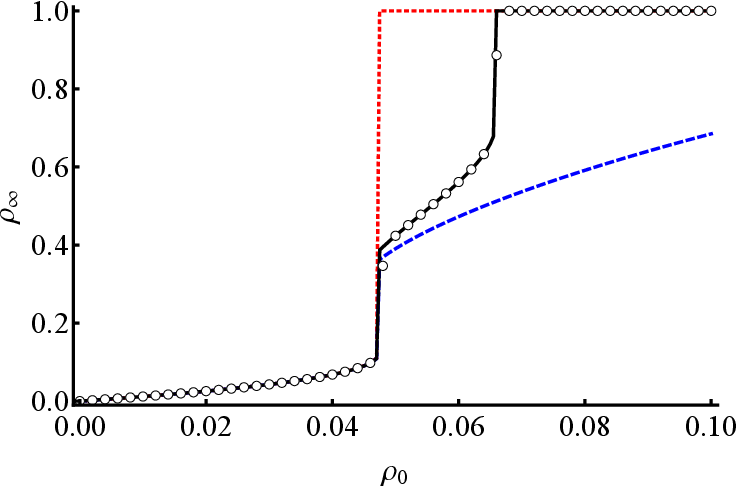}
(b)
\includegraphics[width=7.5cm]{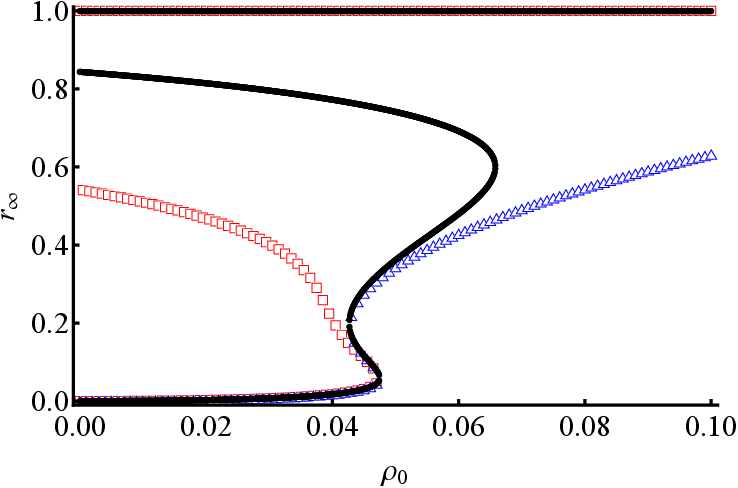}
\caption{
(a) Active node fraction $\rho_\infty$ and (b) fixed points $r_\infty$ of Eq.~(\ref{eq:r}) for the extended Watts model on an ERRG with $\langle k \rangle = 10$.
The adoption threshold of seed neighbors is set to $\theta_\ea=0.25$.
In (a), the red dotted line, black solid line, and blue dashed line represent $\rho_\infty$ when $\theta_\ma=0.50$, $0.67$, and $0.90$, respectively.
In (b), the red open squares, black filled circles, and blue open triangles represent $r_\infty$ when  $\theta_\ma=0.50$, $0.67$, and $0.90$, respectively.
The open circles in (a) represent the simulation results for ERRGs with $10^5$ nodes, averaged over $10^3$ samples ($10^2$ graphs $\times$ $10$ initial states).
The simulation results are in good agreement with the theoretical line, exhibiting two jumps of $\rho_\infty$.
}
 \label{fig:doubleTippingPoints}
 \end{figure}

\begin{figure}[tb]
\centering
(a)
\includegraphics[width=5.2cm]{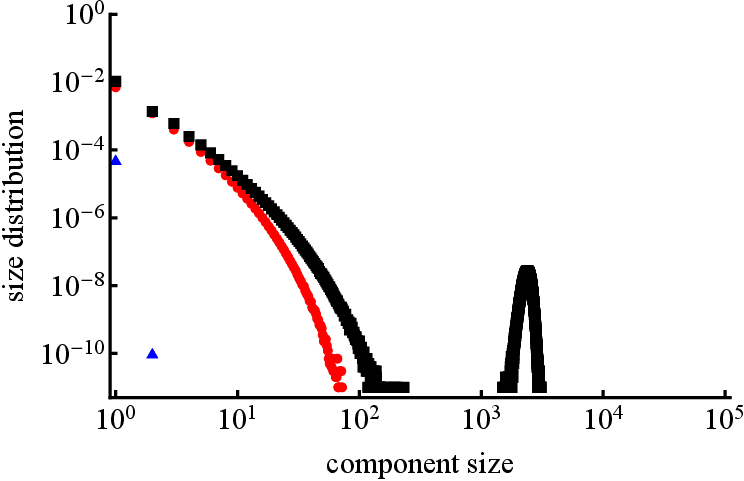}
(b)
\includegraphics[width=5.2cm]{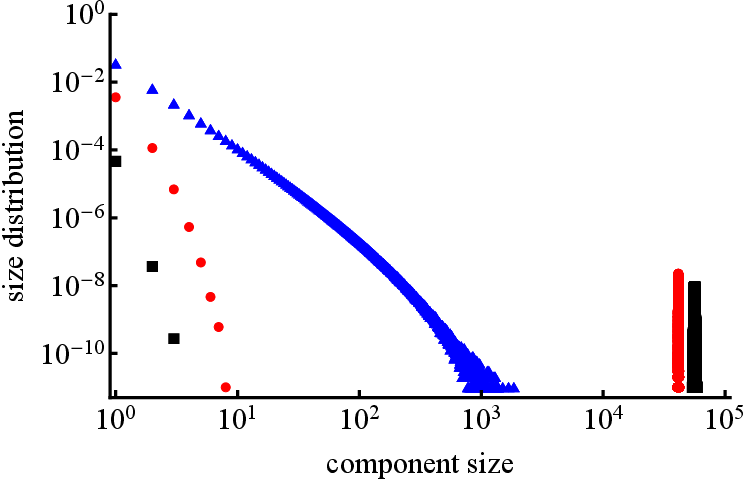}
(c)
\includegraphics[width=5.2cm]{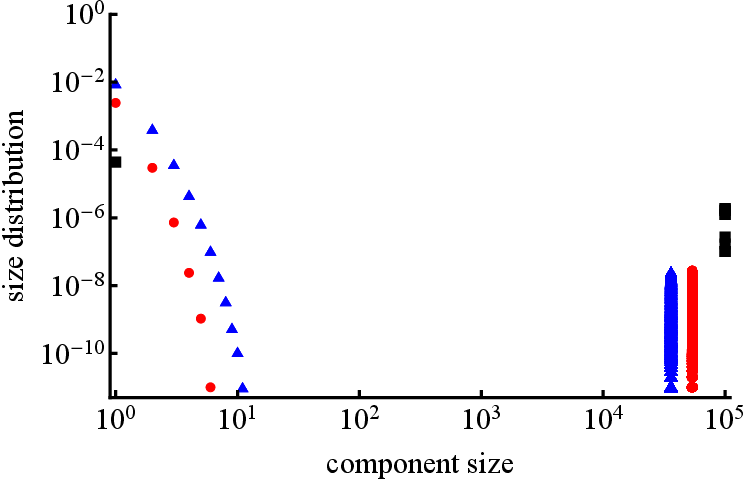}
\caption{
Size distribution of active components (black squares), active SN-components (red circles), and active ON-components (blue triangles) for (a) $\rho_0=0.03<\rC^{\rm 1st}$ (below the first tipping point), (b) $\rC^{\rm 1st}<\rho_0=0.06<\rC^{\rm 2nd}$ (above the first tipping point and below the second tipping point), and (c) $\rho_0=0.09>\rC^{\rm 2nd}$ (above the second tipping point). 
The adoption thresholds of seed neighbors and ordinary nodes are set to $\theta_\ea=0.25$ and $\theta_\ma=0.67$, respectively.
Each distribution, defined as the number of corresponding components of size $s$ per node, is obtained from Monte Carlo simulations of $10^6$ samples ($10^4$ initial states $\times$ $10^2$ graphs) for the extended Watts model on ERRGs with $10^5$ nodes.
}
 \label{fig:activeComponents}
 \end{figure}

We further consider the case where $\theta_{\rm SN} < \theta_{\rm ON}$, meaning that seed neighbors are more easily activated than ordinary nodes due to their proximity to seeds (the case where $\theta_{\rm SN} > \theta_{\rm ON}$ is discussed in Appendix~\ref{sec:AppendixB}).
The coexistence of a giant SN-component and giant ON-component suggests that a large cascade on each giant component can occur at different tipping points. 
Here, we demonstrate that for the extended Watts model, a tipping point exhibiting a jump in $\rho_\infty$ can appear twice.
Let us consider the extended Watts model on an ERRG with $\langle k \rangle = 10$, where a giant SN-component and giant ON-component coexist for $0.011 \lessapprox \rho_0 \lessapprox 0.207$.
Figure~\ref{fig:doubleTippingPoints}(a) illustrates the active node fraction $\rho_\infty$ when $\theta_\ea=0.25$ and $\theta_\ma = 0.50, 0.67$, and $0.90$. 
When $\theta_\ma=0.67$ (i.e., there is a gap between the adoption thresholds of seed neighbors and ordinary nodes), there are five fixed points for $0.043 \lessapprox \rho_0 \lessapprox 0.047$, three fixed points for $0.047 \lessapprox \rho_0 \lessapprox 0.066$, and one fixed point for $\rho_0 \gtrapprox 0.066$ (Fig.~\ref{fig:doubleTippingPoints}(b)). 
Consequently, the active node fraction exhibits a jump at $\rho_0 = \rC^{\rm 1st} \approx 0.047$ and at $\rho_0 =\rC^{\rm 2nd} \approx 0.066$ as $\rho_0$ increases.

A natural interpretation of the existence of two tipping points is that the first cascade at $\rho_0 = \rC^{\rm 1st}$ occurs only on the giant SN-component, while the second cascade at $\rho_0 = \rC^{\rm 2nd}$ spreads to the giant ON-component.
Figures~\ref{fig:activeComponents}(a)--(c) illustrate the size distributions of connected components of active nodes (active components), connected components of active seed neighbors (active SN-components), and connected components of active ordinary nodes (active ON-components) at $\rho_0=0.03$ ($<\rC^{\rm 1st}$), $\rho_0=0.06$ ($>\rC^{\rm 1st}$ and $<\rC^{\rm 2nd}$), and $\rho_0=0.09$ ($>\rC^{\rm 2nd}$).
At $\rho_0 = 0.03$, the size distribution of the active components becomes bimodal, indicating that a giant component is formed by active nodes (Fig.~\ref{fig:activeComponents}(a)). 
However, the giant component observed here is composed of the seeds (which are always active) and the seed neighbors activated by the seeds, and does not indicate that activations propagated.
Figure~\ref{fig:activeComponents}(a) shows that neither the size distribution of the active SN-components nor that of the active ON-components is bimodal.
Above the first tipping point, the bimodality of the active component distribution suddenly becomes more pronounced, indicating the occurrence of large cascades (Fig.~\ref{fig:activeComponents}(b)).
Large cascades for $\rC^{\rm 1st} < \rho_0 < \rC^{\rm 2nd}$ result from the propagation of activations on the giant SN-component.
In fact, the size distribution of the active SN-components is bimodal, whereas that of the active ON-components is not (Fig.~\ref{fig:activeComponents}(b)).
Above the second tipping point, the size distribution of the active ON-components shifts to a bimodal distribution (Fig.~\ref{fig:activeComponents}(c)), signifying that activations can spread across the giant ON-component.

\section{Summary}

\begin{figure}[tb]
\centering
\includegraphics[width=8cm]{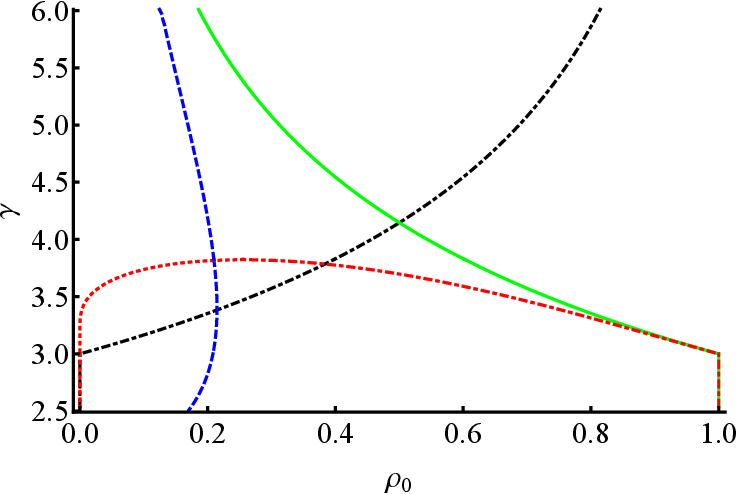}
\caption{
Critical point $\rS$ for S-components (black dot-dashed line), $\rD$ for SN-components (red dotted line), $\rDI$ for (SN+ON)-components (green solid line), and $\rI$ for ON-components (blue dashed line) in the $(\rho_0, \gamma)$ plane for random scale-free networks with $p_k = k^{-\gamma}/\sum_{k'=k_{\rm min}}^{\infty} k'^{-\gamma}$. 
Here we set $k_{\rm min}=2$.
Two critical points for SN-components exist when $\gamma<\gamma_c \approx 3.825$.
}
 \label{fig:scaleFreeNetwork}
 \end{figure}

In this study, we investigated the impact of initiators (seeds) on the tipping point in the cascade dynamics of the Watts model in networks.
To this end, we employed the extended Watts model, where seed neighbors (nodes with connections to initiators) have a different adoption threshold than ordinary nodes (nodes without connections to initiators).
By applying the tree approximation to the extended Watts model on ERRGs, we determined the cascade sizes for various combinations of adoption thresholds for seed neighbors and ordinary nodes.
The results indicate that the tipping point, representing a sudden increase in the active node fraction in the final state, depends on two factors: whether a giant SN-component is formed when the seed fraction is large enough to trigger cascades among seed neighbors, and whether the giant ON-component is maintained when newly activated nodes trigger further activations among ordinary nodes.
We derived the size of the giant component formed by each type of node and its associated critical point, demonstrating how giant components appear or disappear with increasing seed fraction.
When the average degree of the ERRG is large ($\langle k \rangle > \langle k \rangle_c \approx 2.580$), both the giant SN-component and the giant ON-component exist within a certain range of $\rho_0$. The coexistence of a giant SN-component and giant ON-component suggests that large cascades on each giant component can occur at different tipping points.
We presented an example in which a tipping point appears twice when there is a gap between the adoption thresholds of seed neighbors and ordinary nodes.
Through the size distribution of connected components of active nodes of different types, we observed that the initial cascade at a smaller tipping point occurs exclusively in the giant SN-component, while the subsequent cascade at a larger tipping point occurs in the giant ON-component as well.

In this study, we focused solely on ERRGs. 
An interesting question is whether the results presented here hold for other network topologies, particularly for scale-free networks. 
The behavior of the extended Watts model on scale-free networks warrants further investigation due to their degree heterogeneity and the significant differences between nodes constituting the SN-components and ON-components under random seed placement. 
We note that the formation of each type of giant component is strongly influenced by the degree heterogeneity. 
Figure \ref{fig:scaleFreeNetwork} plots the critical points for each type of giant component in the $(\rho_0, \gamma)$ plane for random scale-free networks with $p_k = k^{-\gamma}/\sum_{k'=k_{\rm min}}^{\infty} k'^{-\gamma}$, obtained using the generating function approach described in Appendix \ref{sec:Appendix}. 
When seeds are placed randomly in scale-free networks, seed neighbors are more likely to include high-degree nodes, while ordinary nodes are less likely to do so. 
Consequently, the giant SN-component easily forms for $\gamma<\gamma_c \approx 3.825$, whereas the giant ON-component disintegrates with an increasing seed fraction. 
It remains unclear and worth further discussion whether a combination of adoption thresholds exists that could result in the emergence of two tipping points in scale-free networks.

As demonstrated in this study, large cascades can occur only in the vicinity of initiators when seed neighbors are well connected to each other.  
To the best of our knowledge, this observation has not yet been reported in previous studies on social contagion processes. 
Increasing the number of initiators (thereby increasing the number of seed neighbors) or increasing the influence of initiators (thereby lowering the adoption threshold of seed neighbors) to facilitate large cascades in the vicinity of initiators may lead to novel strategies for viral marketing. 
Further research is necessary to explore the impact of the influence of initiators on information cascades in networks. 

\section*{Acknowledgment}
This work was supported by JSPS KAKENHI Grant Numbers JP19K03648 and JP21H03425.

\appendix

\section{Derivation of the size of giant components formed by nodes of different types and their associated critical points} \label{sec:Appendix}

In this appendix, we derive the normalized size of giant components formed by nodes of different types and their associated critical points in the extended Watts model with random seed placement. 
Let us consider an infinitely large random network with degree distribution $p_k$.
Now, a fraction $\rho_0$ of nodes are randomly selected as seeds. 
Each node that is not a seed is called a seed neighbor if it is connected to one or more seeds and called an ordinary node if not.
Consequently, the following types of connected components are possible: S-components (connected components consisting of seeds), SN-components (connected components consisting of seed neighbors), ON-components (connected components consisting of ordinary nodes), (SN+ON)-components (connected components consisting of seed neighbors and ordinary nodes), and (S+SN)-components (connected components consisting of seeds and seed neighbors). 
For all types except SN-components, the size of giant components and their associated critical points are already known.

The S-components and (SN+ON)-components correspond to the connected components of occupied nodes and those of unoccupied nodes, respectively, for site percolation on the network~\cite{callaway2000network,li2021percolation}.
The generating functions for the degree distribution $p_k$ and excess degree distribution $q_k = (k+1)p_{k+1}/\langle k \rangle$ are defined as 
\be
G_0(x) = \sum_{k} p_k x^k\;\; {\rm and} \;\; G_1(x) = \sum_k q_k x^k,
\ee
respectively.
When a fraction $\rho_0$ of nodes are randomly selected as seeds, the normalized size of the giant S-component, $S_{\rm S}$, is 
\be
\sS = \rho_0 (1-G_0(\uS)), \label{eq:sS}
\ee
where $\uS$, defined as the probability that a node traversed by a randomly selected edge does not belong to the giant S-component, is given by the following equation:
\be
\uS = \trho + \rho_0 G_1(\uS), \label{eq:uS}
\ee
where $\trho = 1- \rho_0$.
Equation~(\ref{eq:uS}) yields the critical point $\rS$, above which the giant S-component appears, as follows:
\be
\rS = \frac{1}{G_1'(1)}.
\ee

The giant (SN+ON)-component size $\sDI$ is obtained by replacing $\rho_0$ and $\trho$ in Eqs.~(\ref{eq:sS}) and (\ref{eq:uS}) as
\be
\sDI = \trho(1-G_0(\uDI)),
\ee
with
\be
\uDI = \rho_0 + \trho G_1(\uDI).
\ee
Here, $\uDI$ denotes the probability that a node followed by a randomly selected edge does not belong to the giant (SN+ON)-component.
The critical point $\rDI$ associated with the giant (SN+ON)-component is expressed as
\be
\rDI = 1- \frac{1}{G_1'(1)}.
\ee

The (S+SN)-components and ON-components correspond to the observable components and unobservable components, respectively, of the observability model on the network~\cite{yang2012network,hasegawa2013observability,allard2014coexistence,yang2016observability,hasegawa2021observability}.
Following~\cite{yang2012network}, we obtain the normalized size $\sSD$ of the giant (S+SN)-component in random networks as 
\be
\sSD = 1-\rho_0G_0(\uSD) -\trho G_0(\wSD)+\trho G_0(\trho \vSD)-\trho G_0(\trho),
\ee
where $\uSD$ represents the probability that for connected pair $(i, j)$, node $j$ does not belong to the giant (S+SN)-component that node $i$ is a seed; 
$\vSD$ represents the probability that node $j$ does not belong to the giant (S+SN)-component given that node $i$ is a seed neighbor and node $j$ is not a seed; 
and $\wSD$ represents the probability that node $j$ does not belong to the giant (S+SN)-component given that node $i$ is a seed neighbor. 
These probabilities are calculated as follows~\cite{yang2012network}:
\begin{eqnarray}
\uSD &=& \rho_0 G_1(\uSD) +\trho G_1(\wSD), \label{eq:usd}\\
\vSD &=& G_1(\trho) +G_1(\wSD) - G_1(\trho \vSD), \label{eq:vsd}\\
\wSD &=& \rho_0 G_1(\uSD) + \trho \vSD. \label{eq:wsd}
\end{eqnarray}
From Eqs.~(\ref{eq:usd}) and (\ref{eq:vsd}), the critical point $\rSD$ for the (S+SN)-components is $\rho_0$, which satisfies the following condition~\cite{yang2012network}:
\be
\trho G_1'(\trho)(1-\rho_0 G_1'(1)-\rho_0 \trho G_1'(1)^2) =G_1'(1)-1.
\ee

The giant ON-component size $\sI$ is obtained by considering the degree distribution and excess degree distribution of the ON-components~\cite{allard2014coexistence,hasegawa2021observability}:
\be
\sI = \trho G_0(\trho) (1-F_0(\uI)), \label{eq:sI}
\ee
where 
\be
\uI = F_1(\uI). \label{eq:uI}
\ee
Here, $F_0(x)$ and $F_1(x)$ represent the generating functions of the degree distribution and excess degree distribution of the ON-components, respectively:
\begin{eqnarray}
F_0(x) &=& \frac{1}{G_0(\trho)}G_0(\trho(G_1(\trho)x+1-G_1(\trho))), \\
F_1(x) &=& \frac{1}{G_1(\trho)}G_1(\trho(G_1(\trho)x+1-G_1(\trho))).
\end{eqnarray}
In Eq.~(\ref{eq:sI}), $\trho G_0(\trho)$ represents the probability that a randomly selected node is an ordinary node. $F_0(\uI)$ represents the probability that none of the neighbors of a node in the ON-components belongs to a giant component, while $1-F_0(\uI)$ represents the probability that at least one of the neighbors of a node in the ON-components belongs to a giant component.
From Eq.~(\ref{eq:uI}), the critical point $\rI$ for the ON-components is given by $\rho_0$, which satisfies the following condition:
\be
F_1'(1)=\trho G_1'(\trho) = 1.
\ee

Finally, we derive the size of the giant SN-component and the corresponding critical point. 
Suppose that a randomly selected node $i$ is not a seed (i.e., it is a seed neighbor or an ordinary node).
We denote the probability that a neighbor of node $i$ is a seed by $\tS$, the probability that it is a seed neighbor by $\tD$, and the probability that it is an ordinary node by $\tI$.
With random seed placement, it is clear that
\be
\tS = \rho_0, \; \tD = \trho(1-G_1(\trho)), \; \tI = \trho G_1(\trho), \label{eq:t3}
\ee
and $\tS+\tD+\tI =1$.
Furthermore, we introduce the probability $\tDF$ that a neighbor of a randomly selected node  $i$ is a seed neighbor and does not belong to a giant SN-component, under the condition that node $i$ is not a seed.
The following equation yields $\tDF$:
\begin{eqnarray}
\tDF
&=& \trho \sum_k q_k \sum_{m=1}^k \binom{k}{m} \tS^m \sum_{\ell=0}^{k-m} \binom{k-m}{\ell} \tI^{k-m-\ell} \tDF^\ell \nonumber \\
&=& \trho \sum_k q_k \sum_{m=1}^k \binom{k}{m} \tS^m (\tI + \tDF)^{k-m} \nonumber \\
&=& \trho \sum_k q_k (\tS+\tI+\tDF)^k - \trho \sum_k q_k (\tI+\tDF)^k\nonumber \\
&=& \trho (G_1(\tS+\tI+\tDF) - G_1(\tI + \tDF)), \label{eq:tpre}
\end{eqnarray}
where $0 \le \tDF \le \tD$.
Substituting Eq.~(\ref{eq:t3}) into Eq.~(\ref{eq:tpre}) yields
\be
\tDF = \trho(G_1(\rho_0 + \trho G_1(\trho)+\tDF)-G_1(\trho G_1(\trho)+\tDF)) = f(\tDF). \label{eq:t}
\ee
Because the giant SN-component size $\sD$ is given as the probability that a randomly selected node is a seed neighbor and not a member of any finite SN-component, we have
\begin{eqnarray}
\sD
&=& \trho \sum_k p_k \sum_{m=1}^k \binom{k}{m} \tS^m \sum_{\ell=0}^{k-m} \binom{k-m}{\ell} \tI^{k-m-\ell} (\tD^\ell - \tDF^\ell) \nonumber \\
&=& \trho \sum_k p_k \sum_{m=1}^k \binom{k}{m} \tS^m (\tI+\tD)^{k-m} 
-\trho \sum_k p_k \sum_{m=1}^k \binom{k}{m} \tS^m (\tI+\tDF)^{k-m} \nonumber  \\
&=& \trho (1-G_0(\tI+\tD)-G_0(\tS+\tI+\tDF)+G_0(\tI+\tDF)) \nonumber \\
&=& \trho (1-G_0(\trho)-G_0(\rho_0 + \trho G_1(\trho)+\tDF)+G_0(\trho G_1(\trho)+\tDF)). \label{eq:sDpre}
\end{eqnarray}
The critical point $\rD$ for the SN-components occurs when the derivative of the right-hand side of Eq.~(\ref{eq:t}) with respect to $\tDF$ is $1$ at $\tDF=\tD$, that is, $f'(\tD)=1$.
The seed fraction satisfying
\be
\trho (G_1'(1)-G_1'(\trho))=1 \label{eq:rD}
\ee
is thus the critical point, $\rD$, for the SN-components.

For an ERRG with average degree $z$, where $G_0(x)=G_1(x)=e^{z(x-1)}$, the critical points are given by the following equations:
\begin{eqnarray}
&& \rS = 1/z, \\
&& \rDI = 1-1/z, \\
&& z(1-\rSD)e^{-z\rSD} (1- z \rSD - z^2 \rSD (1-\rSD)) = z-1, \\
&& z(1-\rI)e^{-z \rI} = 1, \\
&& z(1-\rD)(1-e^{-z \rD}) = 1. \label{eq:rD2}
\end{eqnarray}
Equation~(\ref{eq:rD2}) for the SN-components has two solutions if $z>z_c=(1+\sqrt{5})/2-\log ((3-\sqrt{5})/2) \approx 2.580$, but no solution if $z<z_c$. 
This signifies that for $z>z_c$, the giant SN-component exists in the region where $\rho_0$ lies between the two values of $\rD$; for $z<z_c$, the giant SN-component does not exist regardless of the value of $\rho_0$.

\section{Case where $\theta_\ea > \theta_\ma$} \label{sec:AppendixB}

\begin{figure}[tb]
\centering
\includegraphics[width=8.cm]{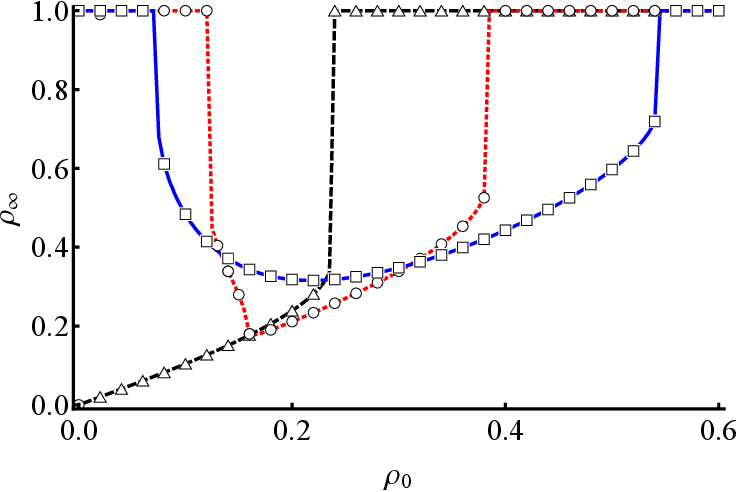}
\caption{
Active node fraction $\rho_\infty$ evaluated from Eq.~(\ref{eq:rho}) and Eq.~(\ref{eq:r}) for the extended Watts model on ERRGs with $\langle k \rangle = 10$.
The black dashed line, red dotted line, and blue solid line represent the results when $(\theta_\ea, \theta_\ma) = (0.5, 0.2)$,  $(\theta_\ea, \theta_\ma) = (0.6, 0.1)$, and $(\theta_\ea, \theta_\ma) = (0.7, 0.0)$, respectively.
The symbols represent the simulation results for ERRGs with $10^6$ nodes, averaged over $10^2$ samples ($10$ graphs $\times$ $10$ initial states).
The theoretical lines are in good agreement with the simulation results.
}
 \label{fig:largeThetaSN}
 \end{figure}

In the main text, we concentrated on the case where $\theta_{\rm SN} < \theta_{\rm ON}$. In this appendix, we address the case where $\theta_{\rm SN} > \theta_{\rm ON}$, indicating that ordinary nodes are more easily activated than seed neighbors. We consider the extended Watts model on ERRGs with $\theta_{\rm SN} > \theta_{\rm ON}$. If activations propagate among seed neighbors, ordinary nodes will also rapidly join the cascade since they are more easily activated than seed neighbors. An interesting scenario arises when $\theta_{\rm SN}$ is significantly high and $\theta_{\rm ON}$ is considerably low. While a significant number of seeds is required for activations to propagate among seed neighbors having a high adoption threshold, a large cascade that spreads across the entire network is more likely to occur when seed fraction $\rho_0$ is small.

Figure~\ref{fig:largeThetaSN} shows the active node fraction $\rho_\infty$ as a function of seed fraction $\rho_0$ for the extended Watts model on an ERRG with $\langle k \rangle =10$, plotted for three different combinations of $\theta_{\rm SN}$ and $\theta_{\rm ON} (<\theta_{\rm SN})$. In all cases, the theoretical lines obtained from the tree approximation (Eq.~(\ref{eq:rho}) and Eq.~(\ref{eq:r})) agree well with the results of the Monte Carlo simulations. The black dashed line for $\theta_{\rm SN} = 0.5$ and $\theta_{\rm ON} = 0.2$ ($\theta_{\rm SN}$ is not too high and $\theta_{\rm ON}$ is not too low) indicates that a large cascade, where $\rho_\infty$ jumps from $\rho_\infty \approx \rho_0$ to $\rho_\infty = 1$, occurs at a certain tipping point $\rho_0 \approx 0.24$ with increasing $\rho_0$.

The interesting case where $\theta_{\rm SN} = 0.6$ and $\theta_{\rm ON} = 0.1$ is illustrated by the red dotted line in Fig.~\ref{fig:largeThetaSN}. We find that $\rho_\infty$ behaves non-monotonically with an increasing seed fraction $\rho_0$: $\rho_\infty = 1$ when $\rho_0$ is small; as $\rho_0$ increases, $\rho_\infty$ drops discontinuously at a certain seed fraction; $\rho_\infty$ decreases and falls to $\rho_\infty \approx \rho_0$ as $\rho_0$ continues to increase; and with further increases in $\rho_0$, $\rho_\infty$ increases again and jumps to $\rho_\infty =1$ at a tipping point.

When $\rho_0$ is small, most nodes are members of the giant ON-component. Even when the activation threshold for seed neighbors is very high, a large cascade that spreads across the entire network is possible if random seed placement leads to a configuration where even a very small number of seed neighbors are surrounded by enough seeds to trigger activation (Monte Carlo simulations sample such a configuration each time when networks are sufficiently large). If $\theta_{\rm ON}$ is sufficiently low, even the activation of a single seed neighbor can trigger a large cascade within the giant ON-component, leading to the further activation of seed neighbors by activated ordinary nodes and resulting in all nodes becoming active.

As the seed fraction increases, the active node fraction drops discontinuously at a certain seed fraction ($\rho_0 \approx 0.13$). This reflects the formation of the giant SN-component and the increased connectivity between seed neighbors, preventing the activation of ordinary nodes from spreading to the giant SN-component. It should be noted that the size of this drop is approximately equal to the fraction of the giant SN-component. As the seed fraction continues to increase, the active node fraction gradually decreases, reaching a minimum at a certain seed fraction ($\rho_0 \approx 0.16$). This roughly coincides with the disappearance of the giant ON-component. In the absence of the giant ON-component, even if some seed neighbors are activated by chance in random seed placement, the propagation of activation to ordinary nodes is negligibly limited. Subsequently, the active node fraction increases continuously with the seed fraction, and jumps to 1 at a tipping point ($\rho_0 \approx 0.39$), indicating the onset of a large cascade on the giant SN-component.

The case where $\theta_{\rm SN} = 0.7$ and $\theta_{\rm ON} = 0.0$ (the blue solid line) is even clearer. In this setting, all ordinary nodes are always active due to their zero adoption threshold. From the figure, we find again that a large cascade which spreads across the entire network occurs for small $\rho_0$ and the active node fraction drops discontinuously at a larger seed fraction ($\rho_0 \approx 0.07$). For the latter, a plausible explanation is that the connection between seed neighbors in the giant SN-component is strengthened, preventing seed neighbors within it from being activated by ordinary nodes (which all become active) and seeds.



\end{document}